\documentclass[12pt,fleqn]{iopart}

\usepackage{iopams}  
\usepackage{verbatim}
\usepackage{xcolor}
\usepackage{graphicx}

\expandafter\let\csname equation*\endcsname\relax
\expandafter\let\csname endequation*\endcsname\relax
\usepackage{mathtools} 

\graphicspath{{images/}}

\DeclareMathOperator\arcosh{arcosh}

\DeclareFontEncoding{FMS}{}{} \DeclareFontSubstitution{FMS}{futm}{m}{n}
\DeclareSymbolFont{hamiltonian-group}{FMS}{futm}{m}{n}
\DeclareMathSymbol{\hamiltonian}{\mathord}{hamiltonian-group}{72} %

\newcommand{\widthParam}{0.95\textwidth}
\newcommand{\widthtTwo}{0.76\textwidth}

\newcommand{\curl}[1]{\nabla\times{#1}}

\renewcommand{\div}[1]{\nabla\cdot{#1}}
\newcommand{\rc}{r_c}
\newcommand{\rh}{r_h}
\newcommand{\heaviside}[1]{\Theta\left({#1}\right)}
\newcommand{\tderivative}[1]{\frac{\rmd{#1}}{\rmd{t}}}
\newcommand{\pderivative}[2]{\frac{\partial{#1}}{\partial{#2}}}

\begin{document}

\title[]{Linear and non-linear instabilities of Kirchhoff's Elliptical Vortices}

\author{Calvin A Fracassi Farias, Renato Pakter, and Yan Levin}

\address{Instituto de F\'{\i}sica, Universidade Federal do Rio Grande do Sul, Porto Alegre, Brasil}
\ead{calvin.farias@ufrgs.br, pakter@ufrgs.br, levin@ufrgs.br}
\vspace{10pt}
\begin{indented}
\item[]June 2020
\end{indented}

\begin{abstract}
We investigate the stability of a uniform elliptical vortex in a two-dimensional incompressible Euler fluid. It is demonstrated that for small eccentricities, the vortex relaxes to a core-halo structure that undergoes rigid rotation with the central core remaining nearly elliptical. For large eccentricities, the vortex splits into two quasi-circular vortices that revolve around the center of mass. Independent of the aspect ratio, the steady-state displays a low-density halo. We present a theory that qualitatively explains the transition between the two states. All theoretical results are compared with extensive molecular dynamics simulations based on the vortex-in-cell algorithm.
\end{abstract}
%
%
%
%

\tableofcontents\newpage

%
\section{Introduction} \label{sc:introduction}
Ever since Onsager's 1949 paper on thermodynamics of turbulence \cite{Onsager1949}, there has been ongoing work on applying the framework of equilibrium statistical mechanics to two-dimensional inviscid fluids \cite{Robert2003, Chavanis1996}. The spontaneous order observed in large-scale vortices, such as Jupiter's great red spot, suggests that there is an underlying fundamental principle that governs the inverse energy cascade \cite{Kraichnan1967, Chen2006}, which results in the formation of large vortex structures. Onsager noted that \(N\) vortices confined to a finite area \(A\) have a finite phase-space volume \(A^N\) and, therefore, the number of accessible microstates, \(\Omega(E)\), for a given energy \(E\) must be a non-monotonic function, such that both for small and large \(E\), \(\Omega(E) \rightarrow 0\). This means that there exists a critical energy \(E_c\) for which \(\Omega(E_c)\) is maximum. The critical energy corresponds to the most disordered configuration of vortices in which they uniformly occupy all the available area. Within the Boltzmann-Gibbs statistical mechanics, the microcanonical entropy is \({S(E) = k_B \ln\Omega(E)}\), which means that for \(E>E_c\), the inhomogeneous vortex distribution will have {\bf lower} entropy than at  \(E_c\). Since the absolute temperature is \({1/T = \partial S(E) / \partial E}\), decreasing entropy entails a state with negative absolute temperatures \cite{Eyink2006}. The existence of negative temperature states implies the clustering of vortices with the same sign of vorticity, which would explain the formation of large vortex structures in the turbulent 2d incompressible flows.

Implicit in Onsager's theory is the ergodicity and mixing of the vortex dynamics on which the thermodynamic argument relies. If Onsager's reasoning is correct, any 2d uniform vortex distribution should relax to a giant circular vortex. On the other hand, it is known from Kirchhoff's work that a uniform elliptic vortex undergoes rigid rotation, maintaining its shape \cite{Kirchhoff1876}.  Furthermore, it has been recently demonstrated that, if perturbed, the Kirchhoff elliptic vortex \cite{Pakter2018} relaxes to a core-halo structure very different from the circular vortex predicted by the Onsager's equilibrium statistical mechanics theory.  

There is an additional subtlety in applying statistical mechanics argument to fluid dynamics. It is well known that Euler equations for an incompressible 2d fluid can be written in terms of vorticity. The vorticity of individual vortices, \(\Gamma_i\), however, must be infinitesimal, so that in the thermodynamic limit when the total number of vortices \(N \rightarrow{} \infty\), the net vorticity \(\Gamma_i N\) remains finite and equal to \(\Gamma_t\). This is precisely the thermodynamic limit for systems with long-range interactions (LRI). It is well known that in the thermodynamic limit, such systems do not relax to the usual Maxwell-Boltzmann equilibrium, but become trapped in non-equilibrium stationary states \cite{Levin2014, Campa2009}. The purpose of this paper is to explore the relaxation into such stationary states of Kirchhoff vortices with different eccentricities.  

\section{Mathematical Formalism of Kirchhoff's Elliptical Vortices} \label{sc:literature-review}
Kirchhoff's vortex is an elliptical patch of  a homogeneous non-viscous fluid of uniform vorticity rotating with constant angular velocity;  it is a particular solution of 2d incompressible Euler equations \cite{Polvani1986}. In the vorticity-stream function formulation, Euler equations can be written as:
\begin{align}
    &\div{\vec{u}} = 0 \label{eq:euler1} \\
    &\pderivative{\Gamma}{t} + \left(\vec{u}\cdot\nabla\right) \Gamma = 0 \label{eq:euler2}
\end{align}
where \(\vec{u} \equiv \vec{u}(\vec{r}, t)\) is the fluid velocity, and \(\vec{r} = (x, y)\) are the coordinates. \(\Gamma \equiv \Gamma(\vec{r}, t)\) is a pseudo-scalar vortex density introduced in the Helmholtz's mathematical formalism of point-vortex model \cite{Helmholtz1867} and defined as \(\Gamma(\vec{r}, t) = \curl{\vec{u}(\vec{r}, t)} \cdot \hat{z}\), where \(\hat{z}\) is the unit vector in the direction normal to the \((x, y)\)-plane. Introducing a stream function \(\psi\) such that  \(\vec{u}(\vec{r}, t) = \curl{\psi(\vec{r}, t)} \hat{z}\) yields the Poisson equation:
\begin{equation} \label{eq:poisson}
    \Delta \psi(\vec{r}, t) = -\Gamma(\vec{r}, t).
\end{equation}
The vortex density can be written in terms of individual vorticity,  \(\Gamma(\vec{r}, t) = \sum_i \Gamma_i \delta(\vec{r} - \vec{r}_i(t))\). Defining the Kirchhoff function as
\begin{equation} \label{eq:hamiltonian}
    \hamiltonian = \frac{1}{2} \sum_{i,j} \Gamma_i \Gamma_j G(\vec{r}_i, \vec{r}_j),
\end{equation}
where \(G(\vec{r}_i, \vec{r}_j)= (-1/2\pi) \ln\left|\vec{r}_i- \vec{r}_j\right|\) is the Green's function solution to the Poisson equation in an open space, the equations of motion for individual vortices take a Hamilton-like form:
\begin{equation} \label{eq:eom}
    \Gamma_i \tderivative{x_i} = \pderivative{\hamiltonian}{y_i}, \qquad\Gamma_i \tderivative{y_i} = - \pderivative{\hamiltonian}{x_i}, \qquad i = 1,2,\dots,N.
\end{equation}
where the coordinates \(x\) and \(y\) are the conjugate variables. Since the Kirchhoff function does not explicitly depend on time, the energy of the system (\(I_3\)) is conserved. Furthermore, it can be shown that the  total vorticity (\(I_1\)), as well as the angular momentum (\(I_2\)), are also conserved quantities: 
\begin{align}
    I_1 &\equiv \int \rmd^2\mathbf{r} \,\Gamma(\vec{r}), \label{eq:total-vorticity} \\
    I_2 &\equiv \int \rmd^2\mathbf{r} \,\Gamma(\vec{r}) \,r^2,  \label{eq:angular-momentum} \\
    I_3 &\equiv \int \rmd^2\mathbf{r} \,\rmd^2\mathbf{r}_0 \,\Gamma(\vec{r}) \,\Gamma(\vec{r}_0) \ln\left|\vec{r} - \vec{r}_0\right|. \label{eq:energy}
\end{align}

\section{Vlasov Dynamics} \label{sc:methodology}
In this paper, we will confine our attention to systems containing $N$ identical vortices of vorticity $\Gamma_i=\Gamma_t/N$, where the total vorticity is
\begin{equation}
    \Gamma_t = \lim_{N \rightarrow{} \infty} \sum_{i=1}^N \Gamma_i \equiv constant.
\end{equation}
Although the pair interaction between any two vortices is vanishingly small, the infinite range of the potential results in a finite total force acting on each vortex. The advection of vorticity by the flow field means that in the thermodynamic limit, \(N \rightarrow{} \infty\), the vortex dynamics is precisely governed by the Vlasov equation \cite{Braun1977}, which for identical vortices is equivalent to Equation \ref{eq:euler2}
\begin{equation} \label{eq:vlasov}
    \pderivative{f}{t} + \pderivative{\psi}{y} \pderivative{f}{x} - \pderivative{\psi}{x} \pderivative{f}{y} = 0,
\end{equation}
where \(f(\vec{r}, t)\) denotes the one-particle distribution function, and \(\psi\) is the mean-field potential (stream function). The one-particle distribution function is related to the vortex density through  \(\Gamma(\vec{r}, t) = \Gamma_t f(\vec{r}, t)\), where we have set the norm of the distribution function to be equal to one. We note, however, that for a finite \(N\) -- as is always the case in computer simulations -- there will exist a residual pair interaction. To avoid these finite-size effects, different methods have been developed, such as the vortex-in-cell algorithm. 

The equations of motion (\ref{eq:eom}) for the vortex dynamics can be written as:
\begin{equation}
    \vec{u}_i = \frac{1}{2\pi} \sum_{j\neq i}^N \frac{\Gamma_j (\vec{r}_i - \vec{r}_j) \times \hat{z}}{r_{ij}^2},
\end{equation}
where \(\vec{r}_i = (x_i, y_i)\) is the coordinate of the \(i\)th vortex, \(r_{ij}\) is the Euclidean distance between the \(i\)th and the \(j\)th vortices, and \(\hat{z}\) is the unit vector normal to the \((x,y)\)-plane. A significant limitation of direct numerical integration of these equations of motion is the order \(O(N^2)\) of the algorithm resulting from the long-range interaction between the vortices. Even if this limitation could be overcome, the finite vortex strength will result in residual pairwise interactions (collisions), which in the large time limit will lead to unreliable results. An alternative is the particle-in-cell/Molecular dynamics simulation(MDS)  algorithm \cite{Dawson1983} developed by \cite{Christiansen1973, Dritschel2008, Dritschel2009}, which allows forces to be calculated using a ``smoothed-out" solution of the Poisson-equation on a grid. The basic steps of the algorithm are the following: \mbox{1) the} space is divided into an \(m\times n\) rectangular mesh and the vortex strength is assigned to each node located in the center of each grid cell; \mbox{2) the} Poisson equation is solved on the grid composed of the nodes using an iterative method; \mbox{3) the} force on each vortex is obtained using the interpolation of the potential obtained in the previous step; \mbox{4) a} 5th order Runge-Kutta algorithm with a controlled step is used to advance the system by a time interval dt; \mbox{5) repeat} until a steady-state is achieved. 
\begin{figure}[!htb]
    \centering
    \includegraphics[width=0.89\textwidth]{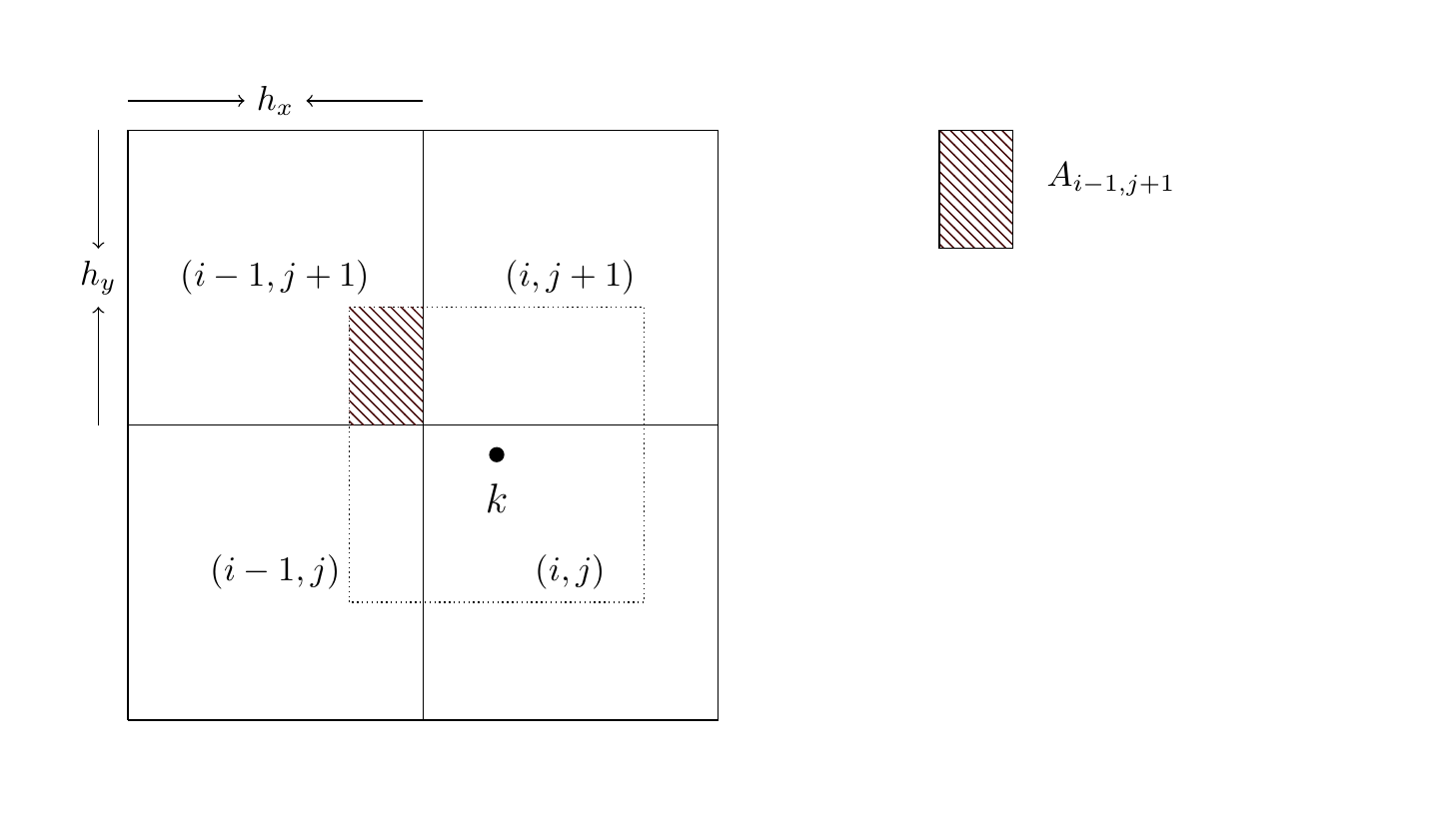}
    \caption{Representation of the mesh-grid that illustrates the particle distribution in the first step of the particle-in-cell algorithm. The \(k\)th vortex, indicated by a black dot, divides its vortex strength between the cell in which it is located and the neighboring cells that are within reach of the PIC kernel, doted square. In the image, the \((i-1, j+1)\)-cell has a contribution proportional to the area \(A_{i-1,j+1}\). The vortex strength is then assigned to the node located in the center of the cell.}
    \label{fig:mesh-grid}
\end{figure}

In the first step, a rectangular mesh with regular nodes at the center of each cell of area \(h_x h_y\) is constructed. Each node is then assigned vorticity in accordance with the Particle-in-Cell algorithm, see Figure \ref{fig:mesh-grid}. Next, the discrete Poisson equation is solved using an iterative technique based on the Gauss-Seidel method, called successive-over-relaxation (SOR) -- both methods are used to solve a linear system of equations of the form \(A \vec{x} = \vec{b}\), but the SOR contains a relaxation factor to accelerate convergence \cite{Grigoryev2002}. Decomposing the matrix \(A\) into a diagonal component \(D\), plus the upper and lower triangular components, respectively \(U\) and \(L\), results in \((D + U + L)\, \vec{x} = \vec{b}\). Then, after a little algebra, it is written as
\begin{equation}
    (D+ \omega L)\, \vec{x} = \omega\vec{b} - \left[\omega U + (\omega -1)D \right]\, \vec{x},
\end{equation}
where \(\omega\) is the relaxation factor constant, which must be greater than or equal to one and less than two. Setting \(\omega\) to unity yields the Gauss-Seidel method. The triangular form, (\(D + \omega L\)), on the left-hand side of the previous equation permits to solve for \(\vec{x}\) iteratively, computing the values \(\vec{x}^{(n)}\) sequentially by forward-substitution, as in the equation below,
\begin{equation} \label{eq:iterative-sor}
    \vec{x}^{(n+1)} = \left.\left.\frac{1}{D + \omega L} \right\{ \omega\vec{b} - \left[ \omega U + (\omega -1)D \right] \vec{x}^{(n)} \right\}.
\end{equation}
The discrete Poisson equation obtained from the density distribution on the mesh-grid in step one is then solved numerically by the above equation, where \(\vec{x}\) represents the unknown potential on the grid nodes and \(\vec{b}\) the mesh-grid density. The matrix \(A \equiv D + U + L\) describes the finite difference approximation of the Laplacian in the Poisson equation according to a 2d five-point stencil, that is, the potential has to be smooth only when considering the first neighbors.
Rewriting Equation \ref{eq:iterative-sor} for each term \(x_i^{(n)}\) yields
\begin{equation}
    x_i^{(n+1)} = (1 - \omega) x_i^{(n)} + \frac{\omega}{a_{ii}} \left(b_i -\sum_{j<i} a_{ij} x_j^{(n+1)} - \sum_{j>i} a_{ij} x_j^{(n)} \right),
\end{equation}
where \(a_{ij}\) are the matrix components, and \(b_i\) are the density components. The estimate of the error of the method is of order of \(O(h^2)\), when using a square grid with \(h = h_x = h_y\).
The complete description of the SOR algorithm can be found in the Ref. \cite{templates}. Interpolating the potential, the force acting on each vortex can be calculated, and the position of the vortices updated using a fifth-order Runge-Kutta algorithm \cite{Ahnert2011}. All the simulations in the present paper were performed with \(N = 2^{20} = 1048576\) point-like vortices, a square grid with \(256\times 256\) cells. The boundary condition is \(\left.\psi(\vec{r})\right|_{\vec{r}\, \in\, \partial A} = 0\) and the potential is continuous at the origin, \(\left. \partial_{\vec{r}} \psi(\vec{r}) \right|_{\vec{r}=0} = 0\).  

\section{Stream function} \label{sc:analysis1}
\begin{figure}[!ht]
    \centering
    \includegraphics[width=0.72\textwidth]{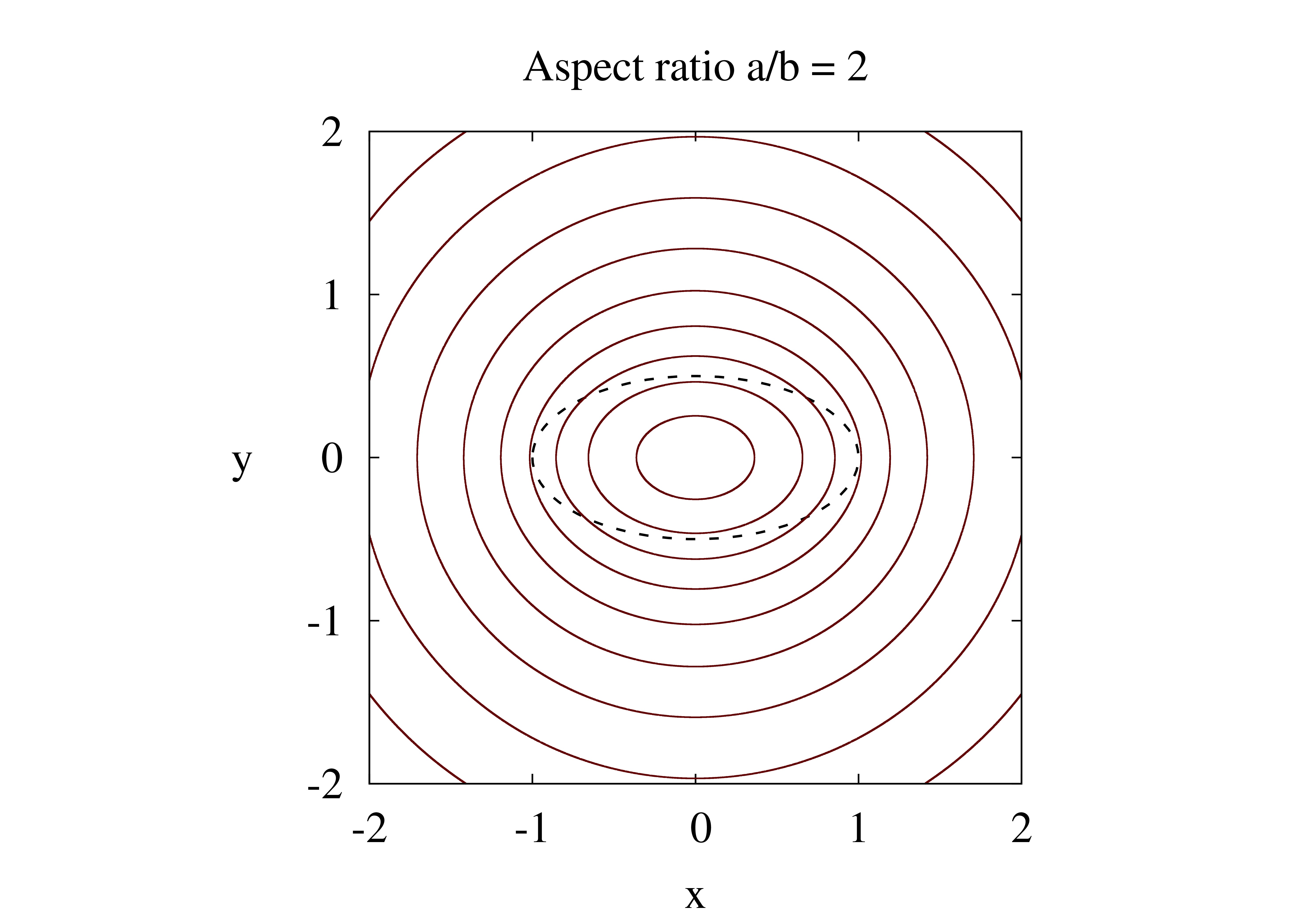}
    \caption{Equipotential lines of the potential of Equation \ref{eq:electrostatic}, \(\psi_\mathrm{ell}(x, y)\) (solid lines). Notice how the border of the distribution function -- dashed ellipse -- crosses the equipotentials. This means that the surface of the ellipse is not an equipotential, signifying that elliptical vortex distribution is not stationary. In this figure, the vortex distribution is a homogeneous ellipse with semiaxes \(a=1.0\) and \(b=0.5\).}
    \label{fig:equipotentials}
\end{figure}

Consider a homogeneous elliptical Kirchhoff vortex, with \(a\) and \(b\), respectively the lengths of the major and minor semiaxes. The stream function produced by the vortex is identical to the electrostatic potential of a two-dimensional ellipse \cite{Kellogg1967, Rizzato2007}.
\begin{equation} \label{eq:electrostatic}
    \psi_\mathrm{ell}(x,y) = \left\{\begin{matrix*}[l] \psi_\mathrm{inn}(x,y), && (x/a)^2 + (y/b)^2 \leq 1 \\ \psi_\mathrm{out}(x,y), && \mathrm{otherwise} \end{matrix*}\right.
\end{equation}
with equations, \(\psi_\mathrm{inn}\) and \(\psi_\mathrm{out}\) given by
\begin{equation} \label{eq:electrostatic_inn}
    \psi_\mathrm{inn}(x, y) = \log\left(\frac{2a}{c}\right) + \frac{1}{2} - \frac{x^2}{a(a+b)} - \frac{y^2}{b(a+b)}  - \arcosh \left( \dfrac{\vphantom{2}a}{c} \right),\, \mathrm{and}
\end{equation}
\begin{equation} \label{eq:electrostatic_out}
    \psi_\mathrm{out}(x,y) = \log\left(\frac{2a}{c}\right) + \frac{1}{2} + \Re\left[\frac{z^2}{c^2}\left(\sqrt{1-\frac{c^2}{z^2}}-1\right)-\arcosh \left( \frac{\vphantom{c^2}z}{c} \right) \right],
\end{equation}
where, \(c \equiv \sqrt{a^2 - b^2}\), \(z \equiv x+\rmi\,y\), \(\rmi = \sqrt{-1}\), and \(\Re\) gives the real part of the argument between the square brackets. Asymptotically, the potential is \(\lim_{r\rightarrow\infty} \psi_\mathrm{ell}(r) = -\log(r)\). As can be seen from Figure \ref{fig:equipotentials}, the boundary of the elliptical distribution crosses various equipotential lines of Equation \ref{eq:electrostatic_inn}. This is also clear from the form of the inner potential in \(\psi_\mathrm{ell}(x,y)\) which has equipotential ellipses with semiaxes proportional to \(\sqrt{a(a+b)}\) and \(\sqrt{b(a+b)}\), while the distribution function has semiaxes \(a\) and \(b\).

To be a stationary solution of the Vlasov equation, the dependence on the coordinates \((x, y)\) in the distribution function \(f(x, y)\) must appear only through the conserved quantities, such as the one particle energy, which in the present case is the stream function \(\psi_\mathrm{ell}\), i.e., \(f \equiv f(\psi_\mathrm{ell}(x,y))\). The fact that the border of the elliptical distribution is not an equipotential surface implies that the original elliptical distribution function is not stationary and will evolve with time.  
The MDS have shown that evolution occurs as a rigid rotation of the ellipse with a constant angular velocity \cite{Pakter2018}. This suggests that a canonical transformation to a rotating reference frame will make elliptical distribution Vlasov-stationary. We, then, consider a rotation,
\begin{align}
    \tilde{x} &= x \cos(\omega t) + y \sin(\omega t), \\ \tilde{y} &= -x \sin(\omega t) + y \cos(\omega t),
\end{align}
the generating function for which is \(F(x, \tilde{y})\), such that \(\tilde{x} = \partial_{\tilde{y}} F(x, \tilde{y})\), and \(y = \partial_{x} F(x, \tilde{y})\). The stream function in the rotating reference frame is \(\widetilde{\psi}_\mathrm{ell}(\tilde{x}, \tilde{y}) = \psi_\mathrm{ell}(\tilde{x}, \tilde{y}) + \partial_t F(x, \tilde{y})\), where \(\partial_t F(x, \tilde{y}) = \omega (\tilde{x} + \tilde{y})/2\). We now look for the angular velocity \(\omega\) which will make the boundary of the ellipse an equipotential of \(\psi_\mathrm{ell}(\tilde{x}, \tilde{y})\). Evaluating the transformed potential at the boundary of the distribution function, i.e., \((\tilde{x}/a)^2 + (\tilde{y}/b)^2 = 1\), shows that it becomes independent of \(\tilde{x}\) and \(\tilde{y}\) for \(\omega = 2/(a + b)^2\). Therefore, the distribution function \(f(\widetilde{\psi}_\mathrm{ell}(\tilde{x}, \tilde{y}))\), with
\begin{equation} \label{eq:elliptical}
    \widetilde{\psi}_\mathrm{ell}(\tilde{x},\tilde{y}) = \psi_\mathrm{ell}(\tilde{x},\tilde{y}) + \frac{\omega}{2}(\tilde{x}^2+\tilde{y}^2), \\ \omega = \frac{2}{(a + b)^2},
\end{equation}
is indeed a stationary solution of the Vlasov equation in the rotating reference \(\mbox{frame \cite{Pakter2018}}\). In the lab frame, the Kirchhoff vortex will then rotate with angular velocity \(\omega = 2 / (a + b)^2\).
\begin{figure}[!ht]
    \centering
    \includegraphics[width=0.76\textwidth]{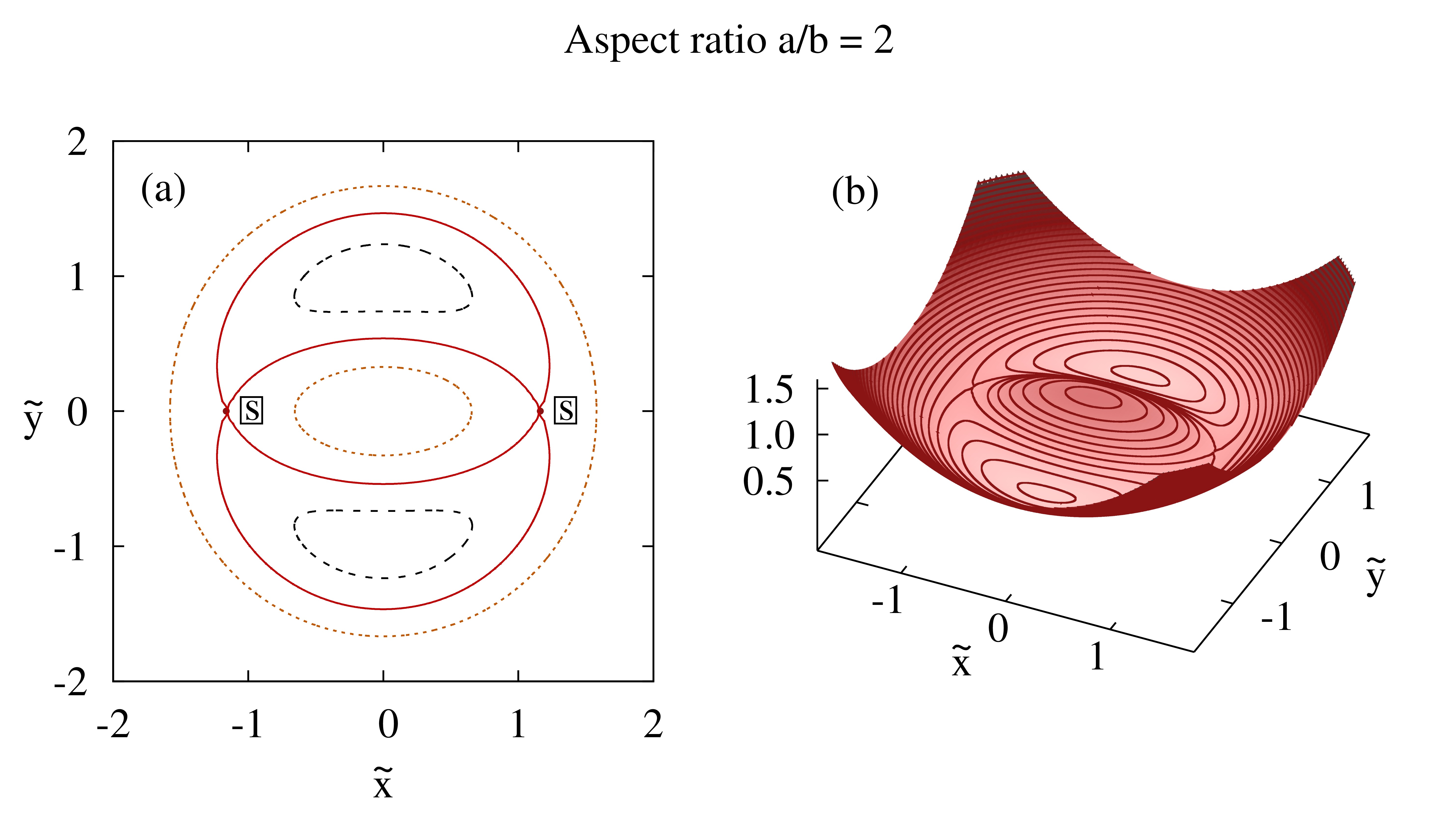}
    \caption{(a) Equipotential curves of the potential in the rotating reference frame, \(\widetilde{\psi}_\mathrm{ell}(\tilde{x}, \tilde{y})\) of Equation \ref{eq:elliptical}, representing the three distinct regions: a separatrix (red solid lines) containing two saddle points (the small dots besides the boxed \(S\)'s), two lower-energy regions (black dashed lines) containing two minima, and two branches of the higher-energy region (orange dotted lines) containing a local maxima at the inner branch. Note that each line color corresponds to the same potential. (b) 3d representation of \(\widetilde{\psi}_\mathrm{ell}(\tilde{x}, \tilde{y})\), notice the contrast between the higher-energy regions (darker shade) to the lower-energy regions (lighter shade).}
    \label{fig:contours}
\end{figure}
Figure \ref{fig:contours}a displays three equipotential curves of the transformed potential, \(\widetilde{\psi}_\mathrm{ell}(\tilde{x},\tilde{y})\), depicted in Figure \ref{fig:contours}b. The red solid line represents the separatrix that separates high and low energy regions of the lab frame; the separatrix contains two fixed hyperbolic (saddle) points located at \((\pm x_\mathrm{fix}, 0)\). Note that in the Figure \ref{fig:contours}a there are different curves which correspond to the {\it same value} of the potential in the rotating reference frame. The equipotentials are organized by the type of line and color -- same line-type (and color) correspond to the same potential energy. The position of the saddle points can be obtained by calculating the derivative of the potential with respect to \(x\), for \(y = 0\), \(\partial_{\tilde{x}}\, \widetilde{\psi}_\mathrm{ell}\left.(\tilde{x},\tilde{y})\right|_{y=0} = 0\). This yields  \(x_\mathrm{fix} = [(a + b)^3 / (a + 3b)]^{1/2}\). The maximum \(y\)-extent, \(y_\mathrm{max}\), along the separatrix trajectory can be computed from the conservation of the potential, \(\widetilde{\psi}_\mathrm{ell}(x_\mathrm{fix}, 0) = \widetilde{\psi}_\mathrm{ell}(0, y_\mathrm{max})\). The black dashed lines represent the low-energy regions that contain two minima located at \((0, \pm y_\mathrm{min})\). The values, \(\pm y_\mathrm{min}\), can be obtained by computing the derivative of the potential with respect to \(y\), for \(x = 0\). Performing the calculation \(\partial_{\tilde{y}}\, \widetilde{\psi}_\mathrm{ell}\left.(\tilde{x}, \tilde{y})\right|_{x=0} = 0\) yields \(y_\mathrm{min} = \, [(a + b)^3 / (3a + b)]^{1/2}\). Finally, the orange dotted lines show the equipotentials corresponding to the same high-energy in the rotating frame. Notice that there are two branches, one inside the separatrix and another outside; they will be henceforth denoted, \(\widetilde{\psi}_\mathrm{ell}^<(\tilde{x},\tilde{y})\) and \(\widetilde{\psi}_\mathrm{ell}^>(\tilde{x},\tilde{y})\), respectively. For further information, we refer the reader to Ref. \cite{Pakter2018}.

\section{Stability of Kirchhoff's vortices} \label{sc:analysis2}
The question of {\it linear} stability of Kirchhoff's vortices is not new; Love's 1893 paper demonstrated that Kirchhoff vortex remains linearly stable as long as the ratio of the major to the minor axes is lower than three \cite{Love1893}. For higher asymmetries, oscillatory surface \(m\) modes will be excited by microscopic perturbations \cite{Burbea1982}. Here, \(m\) represents the number of wavelengths to the elliptic circumference -- similar to Kelvin \(m\) waves that propagate on a circular patch with \(r\cos(m\theta)\) \cite{Thomson1880}. The surface modes of oscillation have dependence only on the elliptical axis ratio and affect the dynamics in different ways; for instance, in the previous section, we showed Kirchhoff's vortices undergo rigid rotation with constant angular velocity. However, rigid rotations only happen for aspect ratios lower than three, which is when the \(m = 3\) mode becomes unstable. As the amplitude of the mode \(m = 3\) grows, the ellipse distorts asymmetrically. On the other hand, when modes \(m = 4\) and \(m = 6\) are dominant, the elliptical patch distorts in a more symmetrical way. We will later see that these modes also affect the final stationary state to which unstable Kirchhoff vortex will relax. The modes \(m = 4\) and \(m = 6\) become unstable at the aspect ratios 4.612 and 7.774, respectively \cite{Mitchell2008}. The \(m = 4\) mode is particularly important. It contains three distinct regimes: filamentation, oscillation, and fission. The fission mode appears at \(a/b = 6.044\) and hints at a new solution, as the nucleus breaks into two blobs of roughly the same size. The early time evolution of the Kirchhoff's vortices were obtained using contour dynamics simulations \cite{Zabusky1979}. Although very accurate, this method can not be used to study the asymptotic stationary states to which the unstable vortices will evolve. For this, we will use vortex-in-cell simulations discussed in Section \ref{sc:methodology}.

\subsection{Linear stability}
We now briefly review the  important properties of the linearly unstable modes \(m = 3\), \(m = 4\), and \(m = 6\). The large time effect of these unstable modes on the asymptotic dynamics of Kirchhoff's vortices will be discussed in the following section.

\begin{figure}[t!hb]
    \centering
    \includegraphics[width=\widthParam]{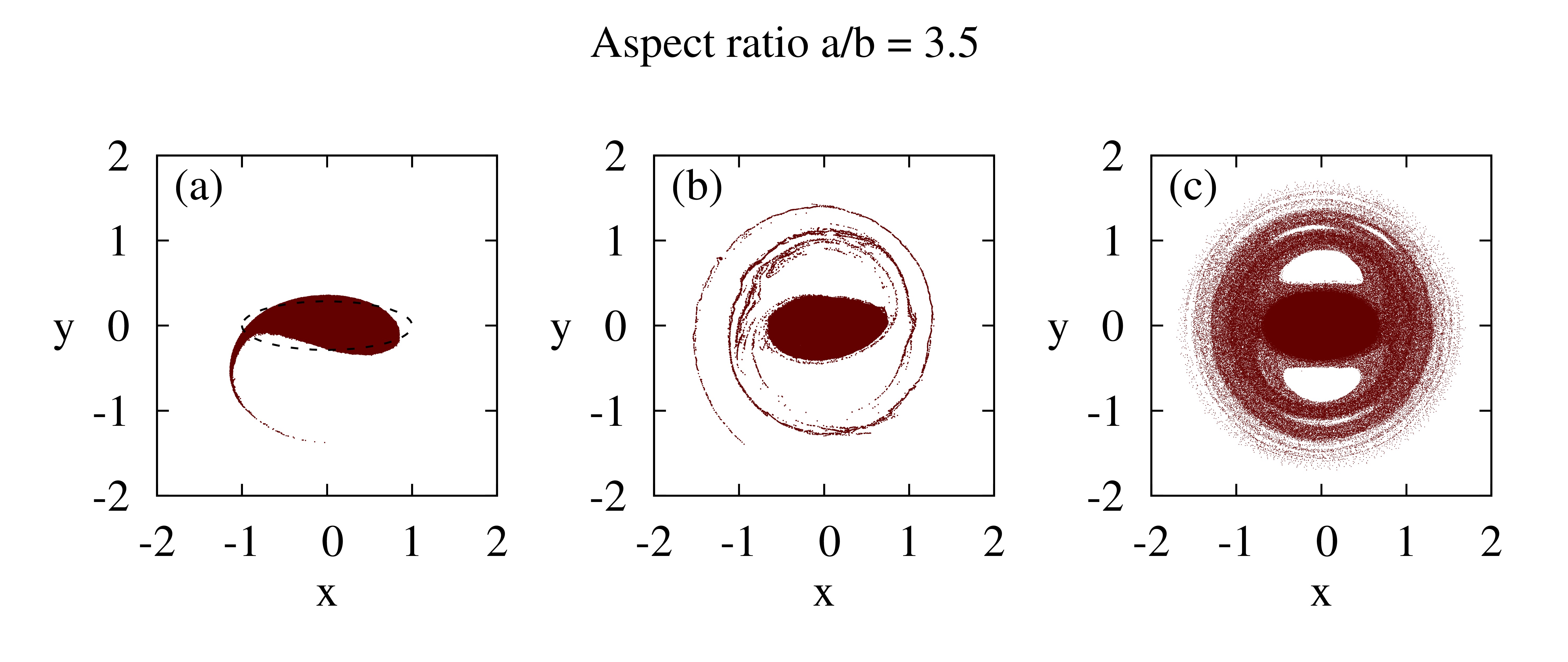}
    \caption{Time evolution snapshots of an ellipse with linearly unstable \(m = 3\) mode (aspect ratio \(a/b = 3.50\)); (a) shows the early evolution of the ellipse against the initial vortex distribution (dashed line); (b) shows the instability filaments taking vortices far beyond the core; and (c) shows the final steady-state with a nearly elliptical core and a thick halo.}
    \label{fig:snap-x3r0350}
\end{figure}
Perturbations become linearly unstable for aspect ratio \(a/b = 3\), with the amplitude of the mode \(m = 3\) growing exponentially. The most notable feature of this unstable mode is that it produces an asymmetric shape, as can be seen in Figure \ref{fig:snap-x3r0350}a; the perturbations lead to a wave propagating on the surface of the ellipse, which eventually breaks \cite{Rizzato2007b}, expelling the vortices away from the main cluster. This leads to the formation of a spiral structure, as can be seen in Figure \ref{fig:snap-x3r0350}b.

\begin{figure}[t!hb]
    \centering
    \includegraphics[width=\widthParam]{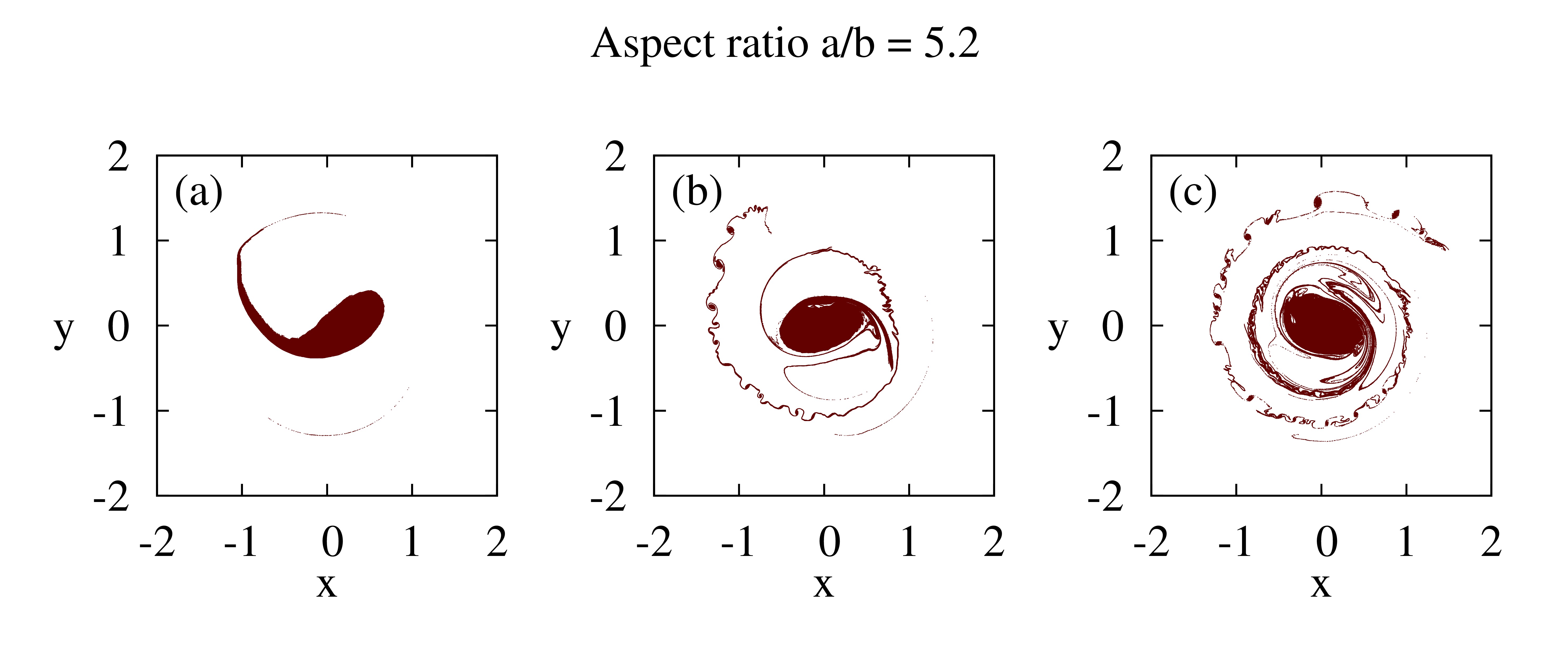}
    \caption{Time evolution snapshots of an ellipse with linearly unstable \(m = 4\) mode (aspect ratio \(a/b = 5.2\)). Note the double spiral, although not symmetrical -- characteristic of the filamentation regime of \(m = 4\) instabilities.}
    \label{fig:snap-x3r0520}
\end{figure}

\begin{figure}[!b]
    \centering
    \includegraphics[width=\widthParam]{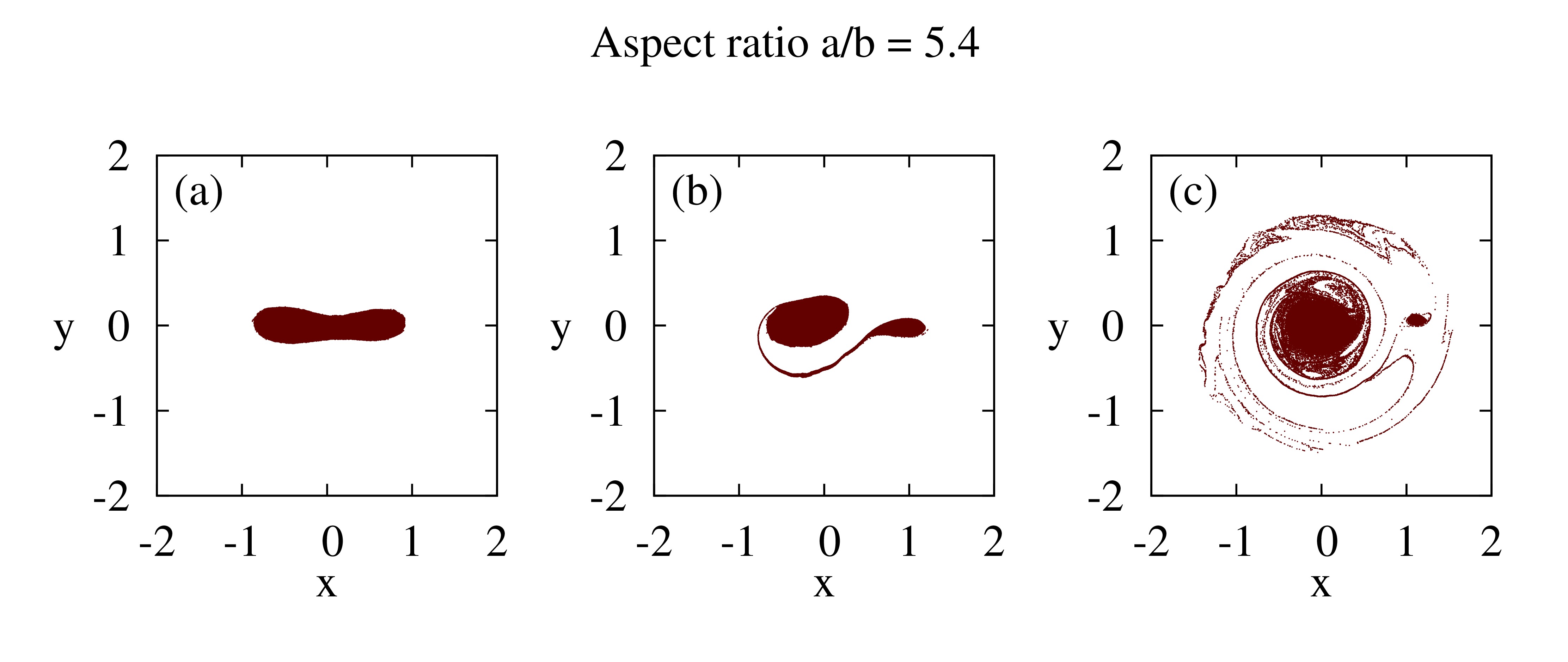}
    \caption{Time evolution snapshots of an ellipse with linearly unstable \(m = 4\) mode (aspect ratio \(a/b = 5.435\)). The evolution illustrates the mechanism by which a dual-core steady-state is formed. Fission must not necessarily occur to form a dual-core structure because, through the filamentation process, the connection between the nuclei becomes so sparse that it forms the halo.}
    \label{fig:snap-x6r0543}
\end{figure}

\begin{figure}[b!th]
    \centering
    \includegraphics[width=\widthParam]{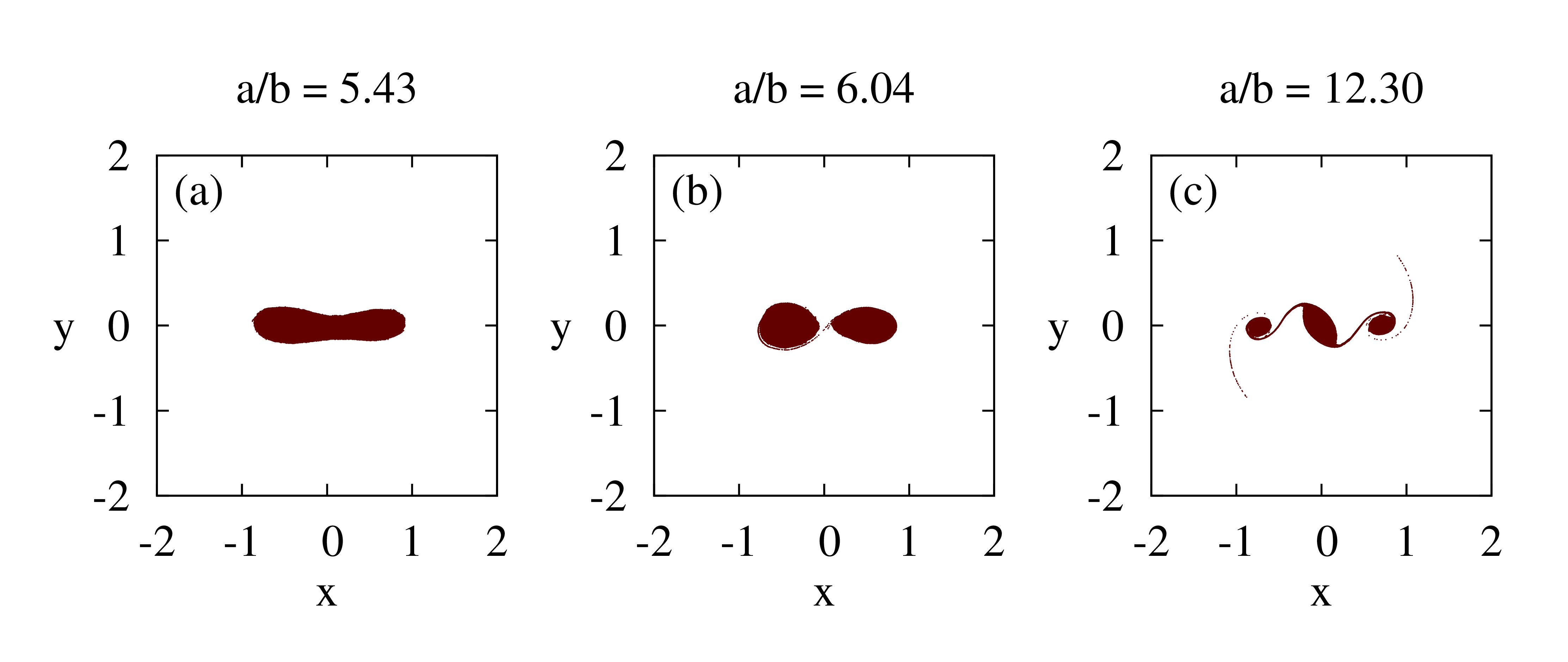}
    \caption{Time evolution snapshots of linearly unstable Kirchhoff's vortices with different unstable modes: (a) \(m = 4\) within oscillation regime; (b) \(m = 4\) within fission regime; and (c) \(m = 6\). The early stages of the evolution show three different clusterings of vortices. Not withstanding the very different short time dynamics, all  these systems evolve to a stationary  core-halo distribution with two asymmetric cores surrounded by halo, revolving around the center of mass.}
    \label{fig:early678}
\end{figure}

The mode \(m = 4\) becomes unstable at the aspect ratio \(a/b = 4.612\) \cite{Mitchell2008}, resulting in a more symmetrical distortion. The first regime is called filamentation and generally behaves as a continuation of the unstable mode \(m = 3\), however presenting a double spiral structure besides the elliptical core, as can be seen in Figure \ref{fig:snap-x3r0520}. The oscillation regime starts at aspect ratio \(a/b = 5.435\) and displays a different evolution. In essence, it results in a
split of the elliptical vortex into two: one large and one small, as can be seen in Figure \ref{fig:snap-x6r0543} and Figure \ref{fig:early678}a. During the subsequent evolution, the small vortex is partially reabsorbed by the large vortex, the process which also results in a significant halo production. Lastly, at aspect ratio \(a/b = 6.044\) the fission regime begins. It is characterized by two identical blobs connected by a thin filament, as can be seen in Figure \ref{fig:early678}b. The filament soon vanishes, resulting in a dual-core structure.

Higher perturbation modes have a very similar dynamics to modes \(m = 3\) and  \(m = 4\). The only remarkable characteristic is the split of the Kirchhoff's elliptical vortex into larger number macroscopic vortices, when the dominant unstable mode has larger \(m \) value, as can be seen in Figure \ref{fig:early678}c, when the dominant mode is \(m = 6\), resulting in two small blobs revolving around the central large blob.

\subsection{Non-linear stability and the core-halo model solution}
\begin{figure}[h!bt]
    \centering
    \includegraphics[width=\widthParam]{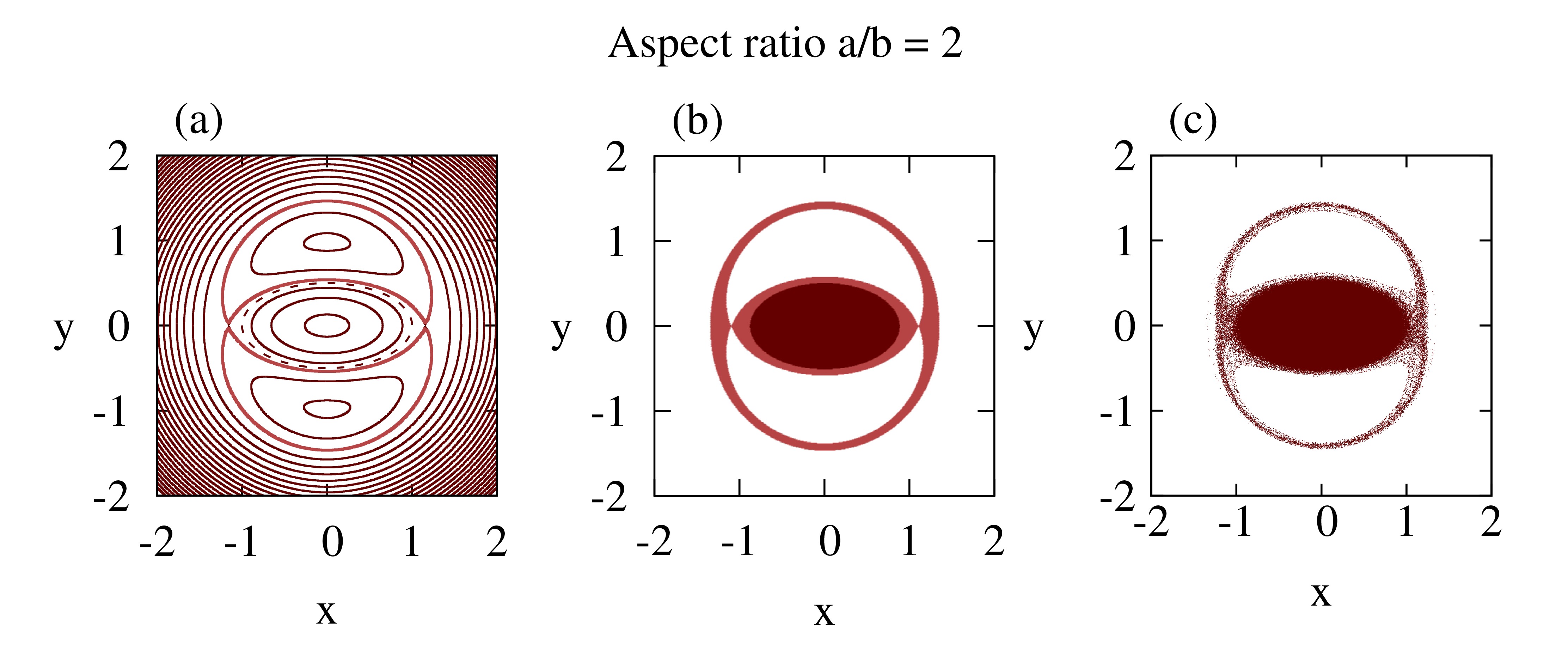}
    \caption{(a) Equipotential curves of the potential in the rotating reference frame, Equation \ref{eq:elliptical}. Notice the resonant structure (thick line) near the original elliptical vortex distribution (dashed line). Perturbed vortices are captured by the separatrix trajectory, resulting  in a thin halo; (b) the theoretical prediction of the ansatz Equation \ref{eq:ansatz}. The core region (darker shade) corresponds to higher density, whereas the halo region (lighter shade) corresponds to lower density; and (c) steady-state of a Kirchhoff elliptic vortex with semiaxes \(a=1\) and \(b=0.5\). 
	}
    \label{fig:contour-x2r0200}
\end{figure}

Molecular dynamics simulations of Kirchhoff's vortices with an aspect ratio of less than three have shown an evolution that differs from the expected behavior. Linear stability analysis predicts that for such aspect ratios, Kirchhoff's vortices should remain stable. This, however, is not what is observed in simulations. Instead, we find that while Kirchhoff vortex rotates, it expels some of the microscopic vortices, resulting in formation of a thin low-density halo. 
Figure \ref{fig:contour-x2r0200}a shows the equipotential curves of the stream function in the rotating reference frame, Equation \ref{eq:elliptical}. On the other hand, Figure \ref{fig:contour-x2r0200}c shows a snapshot of the final steady-state achieved by the Kirchhoff vortex with the aspect ratio \(a/b = 2\). Even though for low eccentricity Kirchhoff vortex is linearly stable, small perturbations can cause non-linear instabilities. In particular, we observe a resonant structure with a separatrix, depicted by the red bold line in Figure \ref{fig:contour-x2r0200}a. Notice that the separatrix is very close to the boundary of the original Kirchhoff vortex. Small perturbation can, therefore, move some of the microscopic vortices outside the main core and allow them to be captured by the separatrix orbit. This resonant mechanism is similar to the non-linear Landau damping in plasma physics \cite{Levin2014, Levin2008, Gluckstern1994}. The halo will form along the separatrix, reaching the maximum radius at around \(y_\mathrm{max}\), corresponding to \(\sim 1.47\) for this specified aspect ratio (illustrated in Figure \ref{fig:contour-x2r0200}). This is consistent with the extent of the halo observed in the simulations.

The formation of halo results in a flow of vorticity from the high-energy to the low-energy regions of the phase space in the lab frame. Since the energy of the whole system is conserved, the removal of some vortices to the low energy region of the phase space (in the lab frame) must be compensated by the clustering of other vortices in the high energy core. The core will then be formed by the process of evaporative heating, which will continue until all the high energy states up to the Fermi energy \(\varepsilon_\mathrm{F}\) in the rotating frame are populated up to the maximum allowed phase space density -- the density of the vortices in the initial distribution. In the rotating frame, the ejected particles will first start to accumulate along the  separatrix of the initial unperturbed ellipse, however, as the core eccentricity decreases due to particle evaporation, the stream function will also change, with the  separatrix orbit moving to energy \(\varepsilon_\mathrm{sep}\).  
The ejection will continue until a stationary state is established, with the rate of ejection from the core equal to the rate of condensation  into the core.  We will then suppose that the halo will be formed by a uniform distribution of density \(\chi\) in: {\it (i)} the region  of the external  branch \(\varepsilon_\mathrm{sep} \le \widetilde{\psi}_\mathrm{ell}^>(\tilde{x}, \tilde{y}) \le  \varepsilon_\mathrm{h}\)  and;  {\it (ii)} the region of the internal branch \(\varepsilon_\mathrm{sep} \le \widetilde{\psi}_\mathrm{ell}^<(\tilde{x}, \tilde{y})\le  \varepsilon_\mathrm{F}\), where \(\varepsilon_\mathrm{sep}\) corresponds to the separatrix energy of the stationary core. The Fermi energy is \(\varepsilon_\mathrm{F} = \widetilde{\psi}_\mathrm{ell}^<(a_s, 0) = \widetilde{\psi}_\mathrm{ell}^<(0, b_s)\), where the self-consistent potential is approximated as that of an ellipse \(\widetilde{\psi}_\mathrm{ell}(\tilde{x}, \tilde{y})\) with semiaxes \(a_s\) and \(b_s\). The separatrix
energy is then \(\varepsilon_\mathrm{sep} = \widetilde{\psi}_\mathrm{ell}(0, y_\mathrm{max})\), where $y_{max}$ is calculated using the semiaxes \(a_s\) and \(b_s\) of the final stationary core. Finally, the halo energy is  \(\varepsilon_\mathrm{h} = \widetilde{\psi}_\mathrm{ell}(0, y_\mathrm{max})\), where $y_\mathrm{max}$ is calculated using the semiaxes \(a\) and \(b\)  of the original Kirchhoff vortex. We, then, propose an ansatz core-halo \mbox{solution \cite{Pakter2018}} to the stationary Vlasov equation in the rotating reference frame: 
\begin{equation} \label{eq:ansatz}
\begin{split}
    f_\mathrm{ch}(\tilde{x}, \tilde{y}) &= \eta\, \heaviside{\widetilde{\psi}_\mathrm{ell}^<(\tilde{x}, \tilde{y}) - \varepsilon_\mathrm{F}} \\ &+ \chi\, \heaviside{\varepsilon_\mathrm{h} - \widetilde{\psi}_\mathrm{ell}^>(\tilde{x}, \tilde{y})} \heaviside{\widetilde{\psi}_\mathrm{ell}(\tilde{x}, \tilde{y}) - \varepsilon_\mathrm{sep}} \heaviside{\varepsilon_\mathrm{F} - \widetilde{\psi}_\mathrm{ell}^<(\tilde{x}, \tilde{y})}
\end{split}
\end{equation}
where $\Theta$ is the Heaviside step function.   Note that the formation of the halo is related only to the non-linear instability produced by the resonant structure of the stream function. The extent of the halo is determined by the \(y_\mathrm{max}\) of the separatrix of the original Kirchhoff vortex, while its width is self-consistently determined by the separatrix of the final elliptical core. 

The stationary distribution parameters \(a_s\), \(b_s\) and \(\chi\) can now be determined using the conservation of the norm of the distribution function, the conservation of the angular momentum, and the conservation of energy, given by Equations (\ref{eq:total-vorticity}--\ref{eq:energy}). For the Kirchhoff vortex with the aspect ratio \(a/b = 2\) we find an excellent agreement between the theory and simulations, see Figure \ref{fig:contour-x2r0200}b and \ref{fig:contour-x2r0200}c. We next explore the domain of validity of the core-halo ansatz solution.

\begin{figure}[!h]
    \centering
    \includegraphics[width=0.75\textwidth]{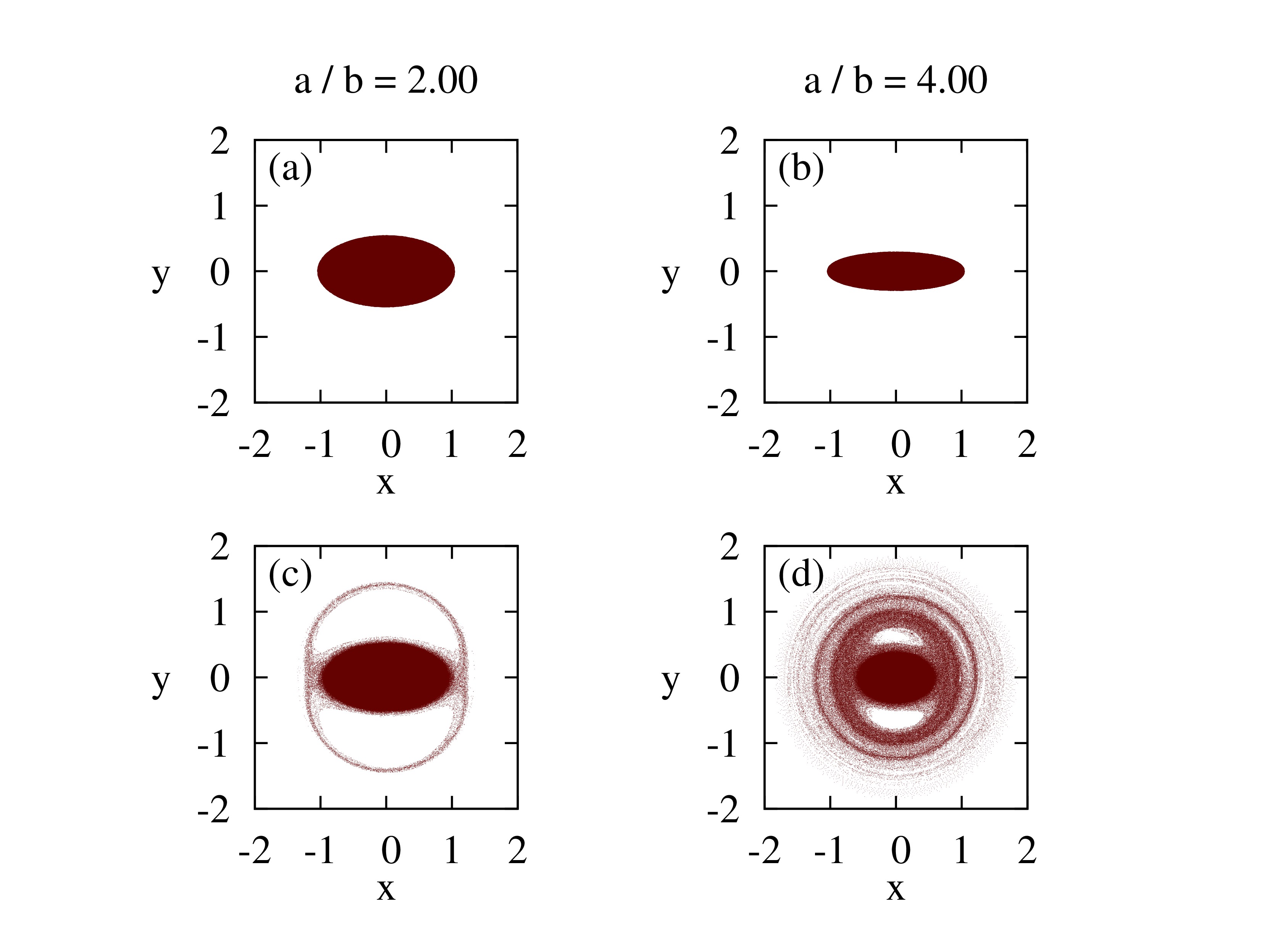}
    \caption{(top row) Snapshots of the initial elliptical distributions of Kirchhoff's vortices with different aspect ratios; and (bottom row) their corresponding stationary states. Notice how evolution leads to a more circular core. The clustering of vortices into the central region (higher energy) compensates for the vortex flow to the halo region (lower energy).}
    \label{fig:proportions}
\end{figure}

\section{Validity of single-core ansatz}
\begin{figure}[!hb]
    \centering
    \includegraphics[width=\widthtTwo]{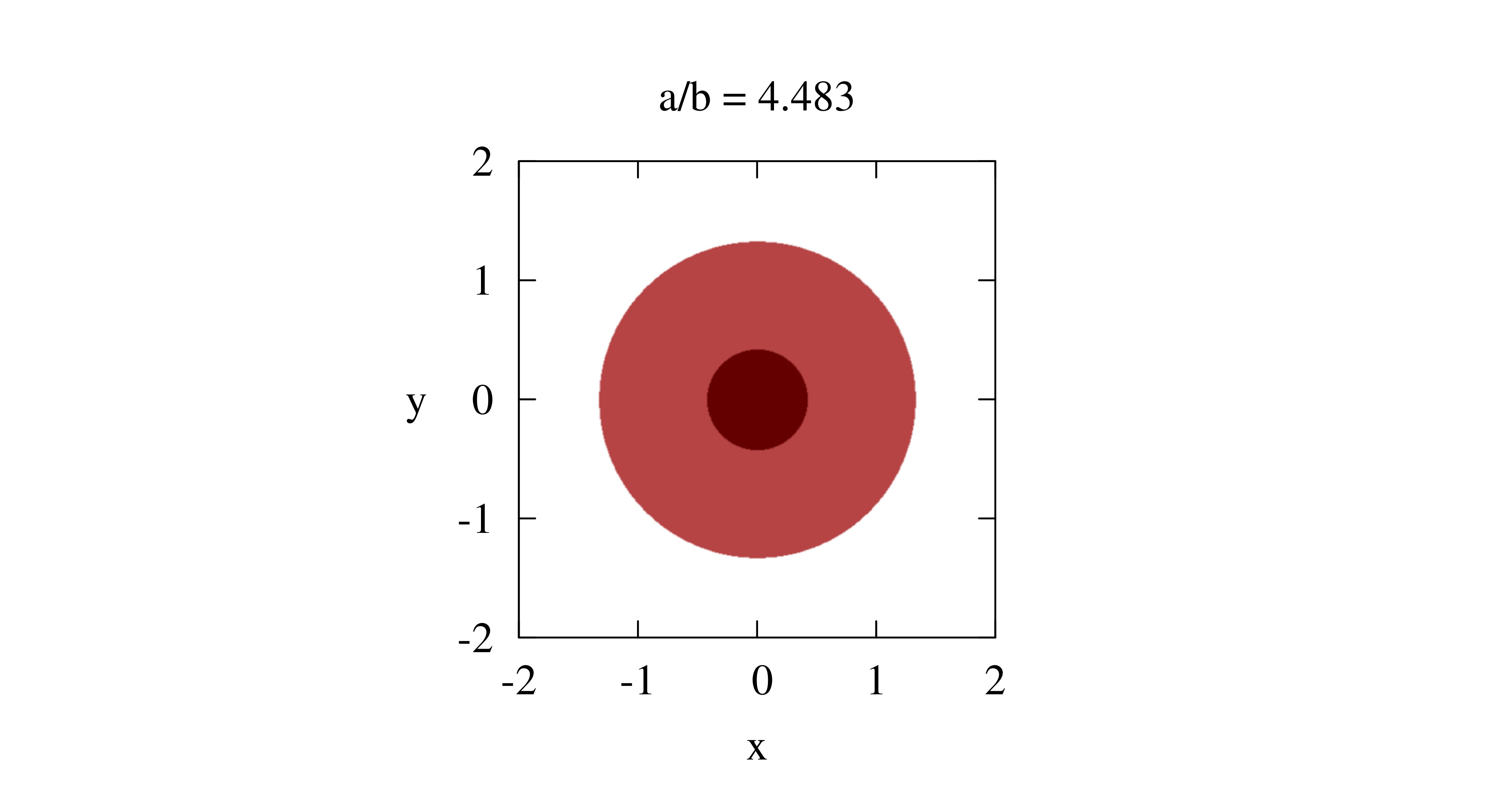}
    \caption{Density distribution of the theoretical circular core-halo distribution, Equation (\ref{eq:circular-ch}), with the parameters determined by solving Equations (\ref{eq:core-radius}), (\ref{eq:conserved-angl}), and (\ref{eq:conserved-energy}). The core region (darker shade) corresponds to the higher density \(\eta \equiv (\pi a b)^{-1}\) (where \(a_c=1, b_c \approx 0.223\)) of the initial distribution, whereas the halo region (lighter shade) corresponds to the lower density \(\chi \approx 0.039\). The core radius is predicted to be \(r_c = 0.424\).}
    \label{fig:snap-XY}
\end{figure}

\begin{figure}[!ht]
    \centering
    \includegraphics[width=\widthParam]{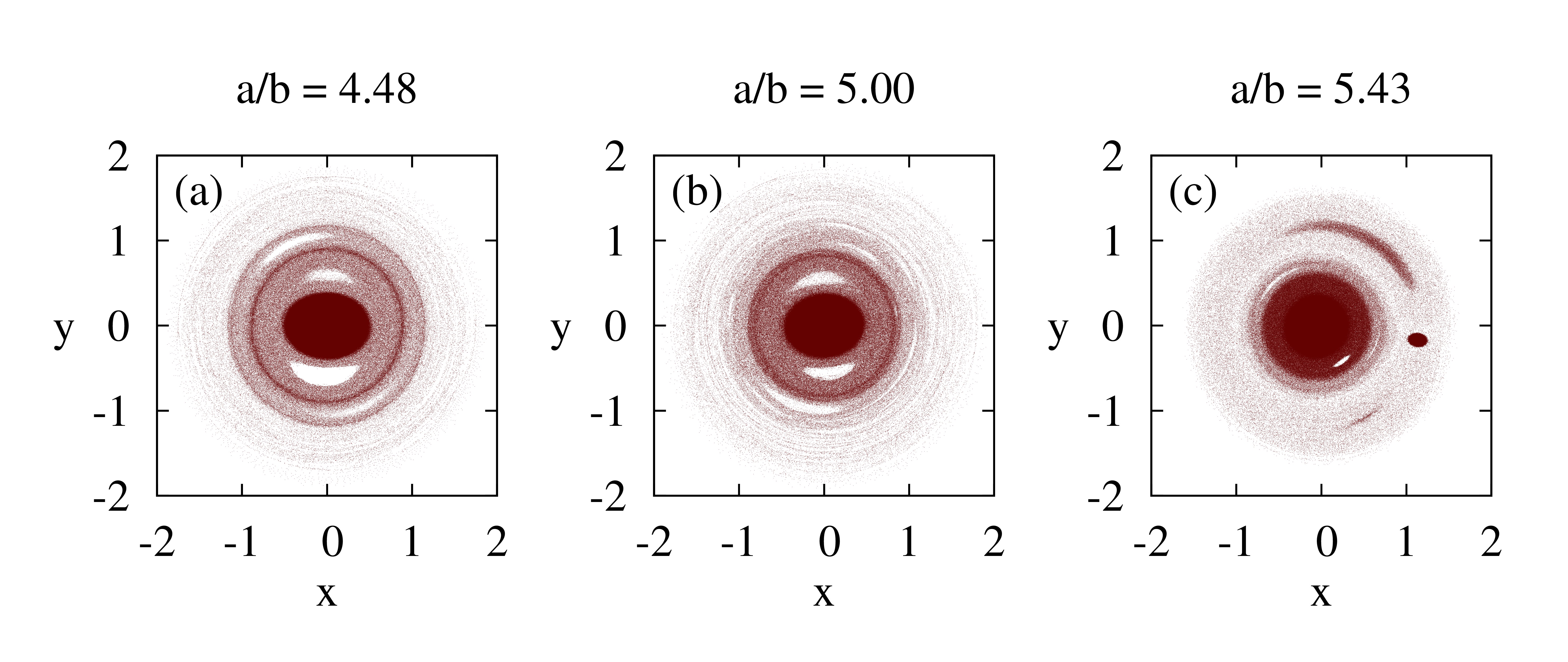}
    \caption{(a) Steady-state for the aspect ratio at which the theory predicts a bifurcation to a new type of solution. Note a more complex halo structure than was assumed in theory. This may explain existence of a single core for this aspect ratio. (b) The core-halo solution with a single core. Note the halo is becoming more uniform. (c) A bifurcation to a new type of solution with two cores.}
    \label{fig:snap-X3m3m4}
\end{figure}

In the previous section, it was shown that a linearly stable Kirchhoff vortex evolves into a core-halo structure. The aspect ratio of the initial Kirchhoff ellipse determines the final steady-state.  We observe, however, that as the eccentricity of the Kirchhoff vortex increases, the halo becomes thicker while the core becomes increasingly circular. This behavior is demonstrated in Figure \ref{fig:proportions}. It is then natural to ask what will happen if the Kirchhoff vortex has eccentricity such that the final stationary core becomes completely circular of radius \(r_c\), while the halo occupies all the area between \(r_c < r < y_\mathrm{max}\), see Figure \ref{fig:snap-XY}. Clearly, beyond this critical value of eccentricity the final stationary state can no longer be of a simple core-halo form and a bifurcation to a new type of solution must occur. To find the critical aspect ratio of the Kirchhoff vortex for which the asymptotic stationary distribution (in the rotating reference frame) must bifurcate to a new form we must determine the aspect ratio \(a_c/b_c\) which will lead to a core-halo distribution with a circular core of radius \(r_c\) and halo of density  \(\chi_c\). Without loss of generality, we can set \(a_c = 1\) and then calculate the critical values \(b_c\), \(r_c\), and \(\chi_c\). A stationary radially symmetric distribution function must have the form \(f(\psi(r))\), which in the case of the core-halo distribution simplifies to 
\begin{equation} \label{eq:circular-ch}
    f_{ch^*}(r) = \eta \heaviside{\rc - r} + \chi_c \heaviside{\rh - r} \heaviside{r - r_c},
\end{equation}
where the halo radius is defined as \(\rh \equiv y_\mathrm{max}\). Then, the solution to the Poisson equation -- whose source function is \(f_{ch^*}(r)\) -- is given by the potential \(\phi(r)\), divided into three distinct regions corresponding to the step levels of the distribution: the potential of the central region (\(\phi_c\)), the potential of the halo region (\(\phi_h\)), and the potential of the outer region, which from the Gauss law is \(-\ln(r)\):
\begin{equation}
    \phi(r) = \left\{\begin{matrix*}[l]
        \quad\phi_c(r), && \,\,0 < r < \rc \\ 
        \quad\phi_h(r), && \rc < r < \rh \\ 
        \,-\ln(r) && \mathrm{otherwise}.
    \end{matrix*}\right.
\end{equation}
Note that the \(-\ln(r)\) behavior of the electrostatic potential for \(r \ge \rh\)  automatically guarantees the conservation of the norm of the distribution function. The stream functions for the core and halo regions are found to be
\begin{align}
	&\phi_c(r) = \left(\pi  \rh^2 \chi_c -1 \right) \log(\rc) - \frac{1}{2} \pi \left[\eta r^2 + \rc^2 (\chi_c -\eta) - \rh^2 \chi_c +2 \rh^2 \chi_c \log(\rh) \right], \\
	&\phi_h(r) = \left(\pi \rh^2 \chi_c -1 \right) \log(r) - \frac{1}{2} \pi \chi_c \left[r^2 - \rh^2 + 2 \rh^2 \log(\rh) \right].
\end{align}
The continuity requirement for the potential's derivative at the interface requires the core radius to be
\begin{align} \label{eq:core-radius}
	&r_c \equiv \left(\frac{\pi \rh^2 \chi_c -1}{\pi \chi_c -\pi  \eta} \right)^{1/2}.
\end{align}
The halo radius \(\rh\) depends on the location of the hyperbolic fixed point corresponding to  the initial vortex distribution, which is only a function of the initial aspect ratio i.e. \(a_c / b_c\). Recall that \(a_c = 1\), so that \(\rh=x_\mathrm{fix}(b_c) = [(1 + b_c)^3 / (1 + 3b_c)]^{1/2}\) and  \(\eta = (\pi b_c)^{-1}\). The critical core radius, therefore,  depends only on \(b_c\) and \(\chi_c\), which can be obtained using the conservation of angular momentum and energy:
\begin{align}
    &\int \rmd^2\mathbf{r}\, r^2 f_{ch^{\textcolor{red}{*}}}(r) = \frac{1}{4} \left( 1 + b_c^2 \right),  \label{eq:conserved-angl} \\
    &\,\frac{1}{2} \int \rmd^2\mathbf{r}\, f_{ch^{\textcolor{red}{*}}}(r)\, \phi(r) = \frac{1}{8} \left[1 - 4 \ln \left( \frac{1+b_c}{2} \right) \right] \label{eq:conserved-energy},
\end{align}
where the right-hand side of the above equations is calculated using the initial Kirchhoff  ellipse with \(a = 1\) and \(b = b_c\). Solving these equations we find \({b_c \approx 0.223}\), \({\chi_c \approx 0.039}\), and \({r_c \approx 0.424}\). The critical aspect ratio beyond which a simple core-halo stationary distribution is no longer possible is found to be \(a_c/b_c\approx 4.48\). 
Observing the results of molecular dynamics simulations we see, however, that for this aspect ratio, the core is still elliptical, see Figure \ref{fig:snap-X3m3m4}a. The bifurcation to a new solution with two asymmetric cores occurs for slightly higher aspect ratio of about \(\sim 5.5\), as can be seen in Figure \ref{fig:snap-X3m3m4}. The error can be attributed to the much more complex halo structure resulting from the overlap of linear and non-linear instabilities. Nonetheless, the simulations are consistent with the theory demonstrating that at the bifurcation point the resulting stationary core is, indeed, circular.  The radius of the stationary core at the bifurcation point predicted by the theory is \(r_c = 0.424\), while the simulations find a slightly smaller circular core of radius \(r_c  \approx  0.35\).

The core-halo model introduced in the previous sections qualitatively explains the steady-state to which a perturbed Kirchhoff vortex will evolve. The model has proven to be quantitatively accurate for Kirchhoff vortices with the aspect ratio less than three, for which linear instabilities are not excited. For linearly unstable Kirchhoff vortices, however, theory is only qualitatively accurate. Wave-breaking  of linearly unstable \(m = 3\) mode leads to the expulsion of microscopic vortices farther than predicted by the resonant structure of the initial Kirchhoff vortex. This can be seen in Figure \ref{fig:snap-x3r0350}a and \ref{fig:snap-x3r0350}b. In addition, the stationary core-halo structure is observed to exhibits a more extensive and thicker halo, as illustrated in Figure \ref{fig:snap-x3r0350}c. Unfortunately, we do not have a theory that allows us to {\it a priori} predict the extent of the halo for Kirchhoff's vortices with \(a/b \ge 3\). The situation becomes even more complex for Kirchhoff vortices with higher eccentricities, when mode \(m = 4\) becomes excited. In this case, the propagating surface waves can lead to fission of the elliptical vortex.

For the aspect ratio \(a/b \approx 5.43\) we begin to observe the asymmetric stationary states with two cores, one of which is much smaller than the other, as seen in Figure \ref{fig:snap-X3m3m4}c. This is consistent with the theoretical argument that a single core state does not exist for Kirchhoff's vortices with sufficiently large eccentricity. The exact value of the aspect ratio at which the bifurcation from a single to double core occurs differs somewhat from the prediction of our theory. This is a consequence of a more complex structure of the halo than was assumed in the theory. 

\begin{figure}[!thb]
    \centering
    \includegraphics[width=\widthtTwo]{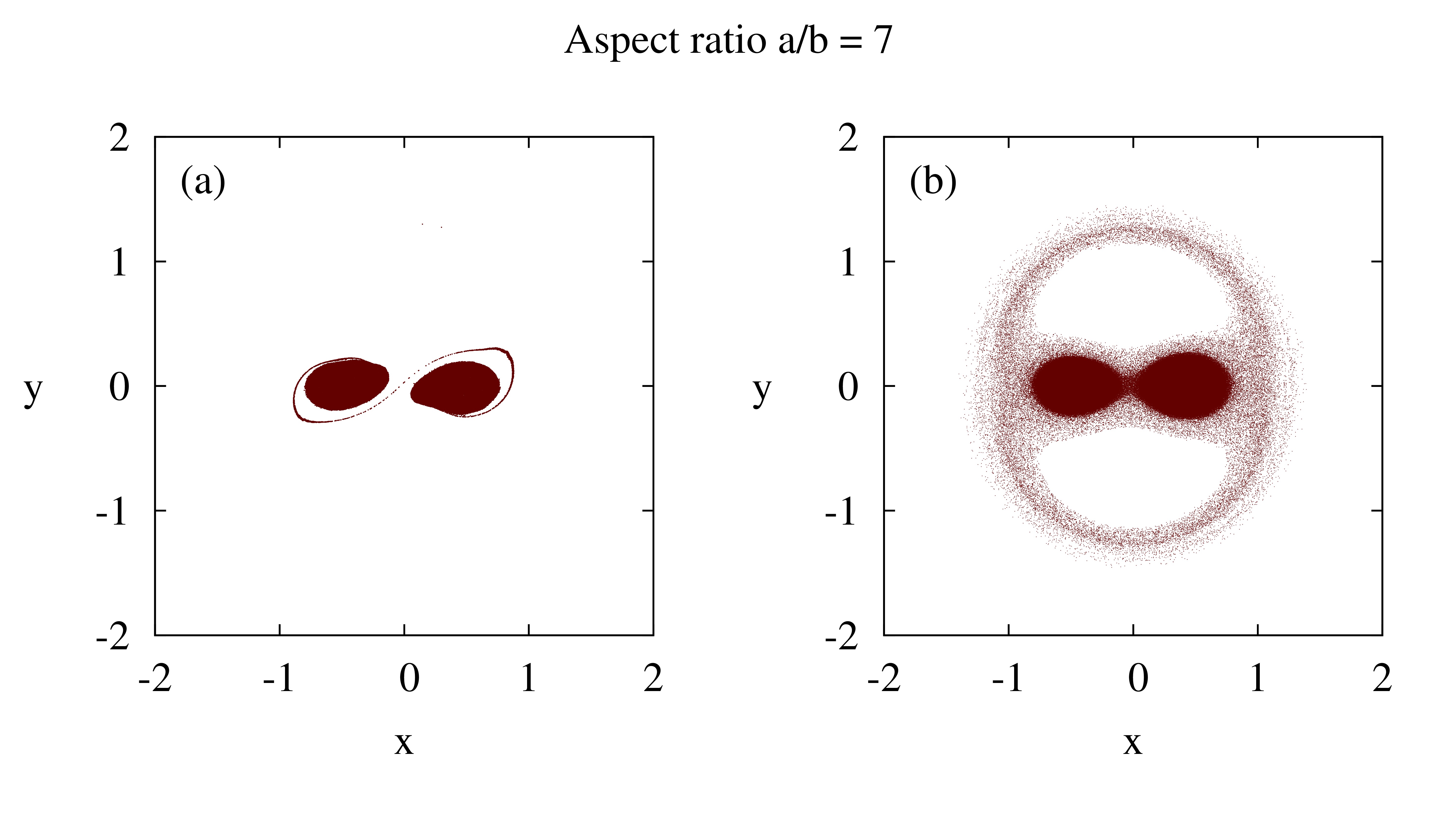}
    \caption{Time evolution snapshots of an unstable Kirchhoff's vortex with aspect ratio \(a/b = 7.00\). Notice the symmetric fission which occurs at the early stages of the evolution, leading to a steady-state with two identical vortices revolving around the center of mass. The halo, once again, is thin.}
    \label{fig:snap-x2r0700}
\end{figure}
For larger aspect ratio,  \(a/b = 7\), the linear instability produced by the \(m = 4\) mode leads to fission of the ellipse into two equal sized cores. The instability happens very rapidly. The subsequent rotation of the two cores around their center of mass, results in a parametric resonance which again expels some of the microscopic vortices from the cores, forming a thin halo, as can be seen in Figure \ref{fig:snap-x2r0700}.

\section{Discussion and Conclusions} \label{sc:discussion}
The core-halo structure is a consequence of non-linear instability, which arises from the resonant structure of the stream function in the rotating reference frame. For an aspect ratio less than three, the Kirchhoff vortex remains linearly stable, and small perturbations are not amplified. The mechanism for the halo formation is the parametric resonance, which moves some of the microscopic vortices into the low energy region of the phase-space. Under these conditions, we can say that the Kirchhoff vortex is linearly stable, but non-linearly unstable. When the eccentricity of the  Kirchhoff vortex increases further, the core becomes linearly unstable. This introduces a new mechanism -- wave-breaking -- for ejecting the microscopic vortices from the core. The halo structure becomes more complex and extends farther than is predicted by the theory. As the aspect ratio increase further, we find that the core becomes more spherically symmetric and the halo more isotropic. Based on this observation, the core-halo model predicts that for large eccentricities, Kirchhoff vortex must relax to a new structure; this is, indeed, found in the simulations, which show that for large eccentricities, the Kirchhoff vortex relaxes to a two-core-halo structure. 

The relaxation of the Kirchhoff vortex into the final stationary state is analogous to the violent relaxation of self-gravitating systems and plasmas. Unlike real particles, however, the individual point vortices of finite strength are physically impossible due to the infinite fluid velocity in the vortex core.  Only vortices of infinitesimal strength are physically relevant and, indeed, the Euler equation can only be mapped onto  a system of  {\bf infinite} number of {\bf infinitesimal}  vortices.  Therefore, unlike gravitational or plasma systems, in which collisional effects eventually lead to the thermodynamic equilibrium after relaxation time which scales with the number of particles, for fluid systems this is not the case -- Euler fluid can only relax to a non-equilibrium stationary state. For this reason we call the steady-state ``stationary",  as opposed to the ``quasi-stationary states" of systems with finite number of real particles interacting through long-range forces. For fluid this is the last stage of evolution, there is no further relaxation. 

To quantitatively study the one-core to two-core transition, we consider an order parameter:
\begin{equation}
    \mathcal{O} = \langle \sin(2\theta) \rangle^2 + \langle \cos(2\theta) \rangle^2,
\end{equation}
where \(\theta\) is the angle that each vortex makes with the \(x\)-axis. An isotropic distribution of vortices would result in an order parameter close to zero, while an elliptical or bimodal particle distribution, such as in the case of a dual-core structure, will lead to a non-zero order parameter.
\begin{figure}[h!t]
    \centering
    \includegraphics[width=0.76\linewidth]{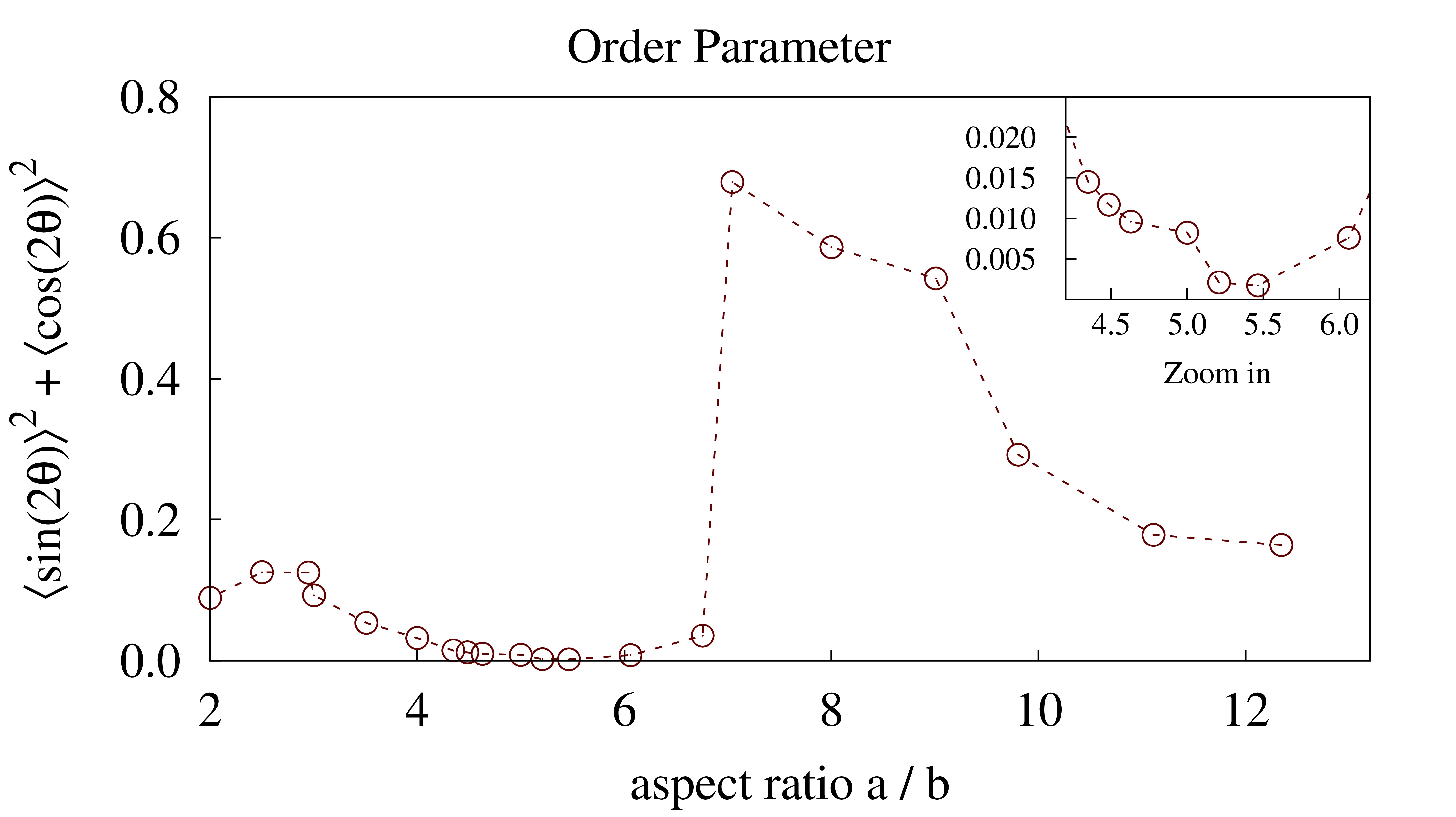}
    \caption{Asymptotic order parameter as a function of the aspect ratio of the initial Kirchhoff vortex. Notice that linearly unstable vortices  \((a/b > 3)\) first result in steady-states which are more isotropic.  However soon after the stationary distribution becomes completely isotropic (\(\mathcal{O} \approx 0\)), there is an abrupt transition to a dual-core structure at \(a/b = 7\).
    }
    \label{fig:orderpar}
\end{figure}
As expected, Kirchhoff vortices with the aspect ratio close to unity result in steady-states with a core-halo structure characterized by a very thin halo and order parameter close to zero. With increasing aspect ratios, the order parameter increases -- corresponding to a higher eccentricity of the core -- up to \(a/b = 3\), where the system is no longer linearly stable resulting in thicker and denser halos. Since halos are close to circularly symmetric, the order parameter decreases. For an aspect ratio between approximately \(a/b = 5.2\) and \(a/b = 5.43\), the particle distribution is almost perfectly isotropic, with the order parameter close to zero, see Figure \ref{fig:orderpar}. For these parameters, the secondary core begins to form. When the aspect ratio reaches the fission values of the \(m = 4\) instability, the order parameter increases dramatically. The \(m = 4\) fission instability, therefore, is responsible for drastically accelerating the formation of the dual-core structure. We would like to stress that the linear-stability analysis by itself is not sufficient to predict the structure of the final stationary state. Even if a Kirchhoff vortex undergoes a fission instability, there is no way to know {\it a priori} that later on one of the cores will not fuse with the other core with an accompanying production of halo. This is indeed what we see happen with Kirchhoff vortices with an aspect ratio close to \(5.5\). On the other hand, the core-halo theory allows us to study the stationary state directly. It predicts that for large aspect ratios there does not exist a stationary (in the rotating reference frame) core-halo solution with a single core.

\section*{Acknowledgments}
This work was partially supported by CNPq, INCT-FCx,
and the US-AFOSR under the grant FA9550-16-1-0280.

\section*{References}
\bibliographystyle{unsrt}
\bibliography{references}

\end{document}